\documentclass[aps,pre,showpacs,twocolumn,amsfonts]{revtex4}
\usepackage{amsmath,pstricks,pst-node,psfig,xy}
\xyoption{all}
\def\pl{\partial}
\def\ot{\leftarrow}
\def\braket#1{\langle {#1} \rangle}

\begin{document}
\title{A Real Space Renormalization Group Approach to Field Evolution
Equations.}
\author{Andreas Degenhard}
\email{andreasd@icr.ac.uk}
\affiliation{The Institute of Cancer Research, Department of
Physics,\\ Downs Road, Sutton, Surrey SM2 5PT, UK}
\author{Javier Rodr\'{\i}guez-Laguna}
\email{javirl@sisifo.imaff.csic.es}
\affiliation{Instituto de Matem\'aticas y F\'{\i}sica Fundamental,
CSIC.\\ C/ Serrano 123, Madrid 28006, Spain.}

\begin{abstract}
A new operator formalism for the reduction of degrees
of freedom in the evolution of discrete partial differential equations
(PDE) via real space Renormalization Group is introduced, in which
cell-overlapping is the key concept. Applications to 1+1-dimensional
PDEs are presented for linear and quadratic equations which are first
order in time.
\end{abstract}

\pacs{02.30.Jr, 02.60.Cb, 05.45.-a, 05.10.Cc 64.60.Ak}
\maketitle


\section{Introduction}
\label{introduction}

The use of Real Space Renormalization Group (RSRG)
techniques~\cite{{ka66},{go93}} to analyze questions related to the
discretization of classical evolution field equations has recently
raised a great deal of attention. Promising results have been achieved
from the concept of {\sl perfect action}~\cite{hn94} and its
application to deterministic partial differential equations (PDE)
\cite{kw97,gkh98}. Recently, the group of Goldenfeld et al. extended
the idea to stochastic PDE \cite{hgk01} by using a space-time
Monte-Carlo formalism for classical problems \cite{zi95}. In this last
work, interesting non-local effects were discovered.

The present work tries to develop further the line traced in
\cite{gkh98} generalizing the notion of {\em coarse-graining}. The fields
are assumed to be defined on spatial cells and a mechanism to define
truncation operators is provided based on the {\em overlapping} of
cells in different partitions of space.
Both linear and non-linear 1+1-PDE are analyzed. Stochastic equations
are not dealt with in the present work, but it should be noticed that
the formalism of \cite{hgk01} may be easily adapted to include the
new truncation operators.

This paper is organized as follows: The next section discusses the
RSRG operator formalism which shall be applied. Our geometric
construction of the truncation operators is explained in detail in the
third section. Section four is devoted to the exposition of some
numerical results. Some concluding remarks and proposals for later
work are discussed in the last section.


\section{The Formalism}
\label{formalism}

Let ${\cal P}$ be a partition of a given region of a manifold ${\cal
M}$, composed of the cells $\{C_i\}_{i=1}^N$. Let $\phi$ be a scalar
field on that region of space and consider the
discretization~\cite{term1} associated to the partition
\begin{align}
\label{partition}
 \phi_i \equiv \int_{\cal M} d\mu\, \phi(x)
\end{align}
where $\mu$ is any measure on ${\cal M}$. Let us, furthermore,
consider the following evolution equation, which we will assume
exact~\cite{term2}:
\begin{align}
\label{exact_evolv}
 \pl_t \phi_i \;=\; H_{ij}\phi_j \;.
\end{align}
This scheme can hold easily any linear evolution equation,
with a great variety of boundary conditions. Equation
(\ref{exact_evolv}) may result from the discretization of any linear
PDE (or even a non-local equation) within any explicit or implicit
algorithm. The operator $H$ shall be termed the {\em evolution
generator}.\\ 
Some non-linear equations may enter easily this
formalism. For example, any quadratic evolution generator might be
added as
\begin{align}
\label{quadr}
 \pl_t \phi_i \;=\; Q_{ijk}\phi_j\phi_k + H_{ij}\phi_j \;.
\end{align}
This allows study of surface growth phenomena as governed by the
Kardar-Parisi-Zhang (KPZ) equation~\cite{kpz86} or the related 1D
turbulence described by Burgers equation. More complex equations such
as Navier-Stokes are by the moment out of reach of the formalism
because the fields under study are not scalar.\\
The field discretizations as defined by equation (\ref{partition})
find their natural place in a vector space $E^N$. A truncation
operator $R:\;E^N\mapsto E^M$ defines a sub-discretization within the
original vector space. The effective field component indices shall be
denoted with capital letters: $\{\phi'_I\}\in E^M$. The new discretization
only provides $M$ degrees of freedom and, thus, the $R$ operator must
have a non-trivial kernel.\\
The truncation operator shall be chosen to be linear~\cite{term3}. 
This enables us to write its action as
\begin{align}
\label{action}
 \phi'_I \;=\; R_{Ii}\phi_i \;.
\end{align}
Had the $R$ operator got a trivial kernel, an inverse
operator might be written, $R^{-1}$, which would be called the
{\sl embedding} operator. In this case the following equation
would be exact
\begin{align}
\label{exact}
\pl_t \phi_i \;=\; H_{ij} R^{-1}_{jJ} \phi'_J \;,
\end{align}
One might therefore evolve the effective discretization with only $M$
degrees of freedom through equation
\begin{align}
\label{discr_evolv}
 \pl_t \phi'_I \;=\; R_{Ii} H_{ij} R^{-1}_{jJ} \phi'_J \;\equiv\;
       H'_{IJ} \phi'_J \;,
\end{align}
where $H'$ is the {\em renormalized evolution
generator}. After the evolution of the reduced discretization has been
completed, the evolution of the original discretization would be
found
\begin{align}
\label{orig_evolv}
\phi_i(t) \;=\; R^{-1}_{iI} \phi'_I(t) \;.
\end{align}
Equation (\ref{discr_evolv}) requires less storage and CPU time than
equation (\ref{exact_evolv}) to be simulated on a computer. We may
express this situation by the commutative diagram
%
\begin{align}
\label{comm_diag}
 \xymatrix{\ar @{} [ddrr] |{=}
           E^N \ar@/_/[dd]_R \ar[rr]^H && E^N \ar@/^/[dd]^R \\ \\
           E^M \ar@/_/[uu]|{R^{-1}} \ar[rr]_{H'} && E^M \ar@/^/[uu]|{R^{-1}}}
\end{align}

Unfortunately, the situation displayed in the previous paragraph is
impossible: the truncation operator must have a non-trivial
kernel. Thus, it lacks a true inverse. Anyway, a ``best possible''
pseudo-inverse may be found: an operator $R^p$ which fulfills the
Moore-Penrose conditions~\cite{gl96}
\begin{align}
\label{moore}
  RR^pR &\;=\;R \qquad\qquad\quad\,  R^pRR^p \;=\;R^p \nonumber\\
   (R^pR)^\dagger &\;=\;R^pR \qquad\qquad (RR^p)^\dagger \;=\;RR^p \quad .
\end{align}
Those equations are solved only if $R^p$ is the singular values
decomposition (SVD) pseudo-inverse of $R$. $R^p$ is an
``extrapolation'' operator, which takes an $E^M$ (reduced)
discretization and returns an approximate $E^N$ (full) one.
The only important piece of information contained in $R$ is its
kernel, which represents the degrees of freedom which are removed
(see, e.g. \cite{gms95}). $RR^p$ is the identity operator on $E^M$ and
$R^pR$ is a projector on the {\em relevant degrees of freedom}
subspace of $E^N$. These degrees of freedom are stored as the column
of the matrix $R$. It is highly recommended to orthonormalize these
column vectors, because $R^p$ becomes simply $R^\dagger$.\\
Using the pseudo-inverse $R^p$ instead of $R^{-1}$ the diagram
(\ref{comm_diag}) does not commute. The ``{\sl curvature}'' represents
the error of the procedure. The renormalized evolution generator is
written as:
\begin{align}
 \label{evolv_gen}
   H'_{IJ} \;=\; R_{Ii} H_{ij} R^p_{jJ}
\end{align}
where indices are kept for clarity. A quadratic evolution generator
would be transformed in this way:
\begin{align}
 \label{renorm_gen}
  Q'_{IJK} \;=\; R_{Ii} Q_{ijk} R^p_{jJ} R^p_{kK} \;.
\end{align}
This expression shall be shorthanded as $Q'=RQR^p$. Higher degree
operators are possible, of course. 

The election of the $R$ operator is the key problem. Ideally it
should depend on the problem at hand, i.e. on the field equation
and the observables we want to measure. In this paper a geometrical
approach is introduced which is independent of the physics of the
dynamical system, but uses a quasi-static truncation procedure
for a careful selection of the relevant degrees of freedom.\\
The schedule for all the simulations that shall be presented in the
rest of this work is:
\vskip 2mm
\noindent%
1.- Present a Hamiltonian $H$ (at most quadratic)
       and an initial field $\phi(0)$.\\[0.1cm]
2.- Perform the exact evolution and obtain $\phi(t)$.\\[0.1cm]
3.- Propose a truncation operator $R$ and obtain
the pseudo-inverse $R^p$.\\[0.1cm]
4.- Calculate the renormalized Hamiltonian and the
truncated initial field: $H'=RHR^p$ and $\phi'(0)=R\phi(0)$.\\[0.1cm]
5.- Perform the renormalized evolution on $\phi'(0)$
and obtain $\phi'(t)$.\\[0.1cm]
6.- Compare $\phi(t)$ and $R^p\phi'(t)$.
\vskip 2mm

We distinguish between a {\em real space error}, which is given by the
$L^2$ norm of $[\phi(t)-R^p\phi'(t)]$ (a vector from $E^N$) and the
{\em renormalized space error}, which is given by the $L^2$ norm of
$[R\phi(t)-\phi'(t)]$ (a vector from $E^M$). Both errors need not be
equal. It is impossible for the first error to vanish for all
$\phi(0)$ and all time, although that is possible for the second
one. In that case, the retained degrees of freedom are {\em exactly
evolved} after the rest of the information has been removed. Such a
situation corresponds to a {\em perfect action}.

\section{Geometric Truncation Operators}
\label{geomtrunc}
\begin{figure}[ht]
\vspace{4mm}
\centerline{
\psfig{figure=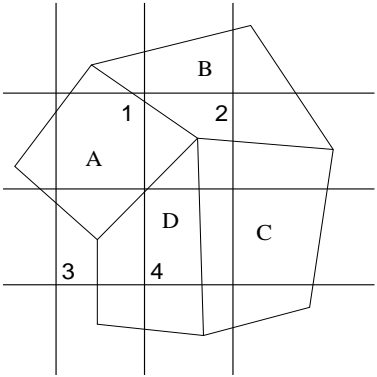,width=5.0cm,height=5.0cm}}
\vspace{0.2cm}
\caption{A part of two overlapping partitions is depicted.
         The lines delimiting the ``old'' partition are thin
         (cells $A$, $B$...), while the thick lines belong to
         the ``new'' one ($1$, $2$,...). For example, there
         shall be no $R$ matrix element between cells $1$ and
         $C$, since they do not overlap. On the other hand,
         the matrix element $R_{1A}$ must be close to $1$.}
  \label{overlap_part}
\end{figure}

In this section a set of construction rules for the $R$ operator
shall be presented which shall allow for practical computations.\\
Let us consider the 1D interval $[0,1]$ and let ${\cal P}_n$ denote a
regular partition of that interval into $n$ equal cells, denoted by
$C^n_i \equiv [\frac{i-1}{n}, \frac{i}{n}]$. The truncation operator
$R^{M\ot N}$ shall be defined by
\begin{align}
 \label{trunc_def}
  R^{M\ot N}_{Ii} \equiv \frac{\mu(C^M_I \cap C^N_i)}{\mu(C^M_I)} 
\end{align}
where $\mu(\cdot)$ denotes the standard measure in $\mathbb R$, $C_i$
is a cell of the source partition ${\cal P}_N$ and $C_I$ is part of
the destination one ${\cal P}_M$. In geometrical terms, the $R$ matrix
elements are given by the ratio
\begin{align}
 \label{overlap}
  R^{M\ot N}_{Ii} \;=\;
    \frac{\hbox{ Overlap between cells $C_i$ and $C_I$.}}
         {\hbox{ Measure of cell $C_I$.}}
\end{align}
The rationale behind this expression may be expressed with a physical
analogy. Let us consider ${\bf \phi}_i$ as the density of a gas in the
$i$-th cell of the source partition, limited by impenetrable
walls. Now a new set of walls is settled: the ones corresponding
to the new (destination) partition. The old walls are, after that,
removed. The gas molecules redistribute uniformly in each new
cell. The new densities are the values ${\bf \phi}_I$ which constitute
the transformed field discretization. Figure \ref{overlap_part} should
be helpful.
%
%
%

In more mathematical terms, the value of ${\bf \phi}_I$ is a linear estimate
for
\begin{align}
 \label{lin_est}
  {\bf \phi}_I \,=\, \int_{C_I} \phi(x) dx 
\end{align}
conserving the total mass: $\sum_I {\bf \phi}_I\,=\,\sum_i {\bf \phi}_i$.
Equation (\ref{trunc_def}) may also remind of the definition for
conditional probability.

\begin{figure}[ht]
\vspace{0.4cm}
\centerline{
  \psfig{figure=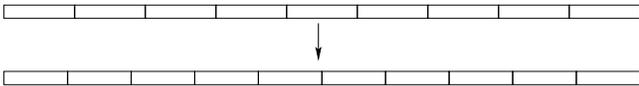,width=8.5cm}}
\vspace{2mm}
\caption{The lower partition has just a single degree of freedom
         less than the one above. A truncation matrix may be
         written to proceed from one to the other.}
  \label{part}
\end{figure}

The resulting $R^{M\ot N}$ operators shall be termed {\em sudden
truncation operators}. Compared to standard RSRG integer factor
blocking techniques~\cite{gkh98}, the operators $R^{M\ot N}$ allow for a
greater flexibility. For example, it is possible to remove a {\em
single degree of freedom} (see figure \ref{part} for a 1D example).
%
%
The sudden truncation operators do not form a closed algebra. The
composition of sudden truncation operators shall take us to the
concept of {\sl quasistatic} or {\sl adiabatic} truncation
operators. These are defined by:
\begin{align}
 \label{adia_trunc}
  qR^{M\ot N} \;=\; R^{M\ot M+1} R^{M+1\ot M+2}
     \cdots R^{N-1\ot N}
\end{align}\\
Of course, $qR^{M\ot N}$ differs greatly from $R^{M\ot N}$. The term
``quasistatic'' is suggested by the thermodynamical analogy introduced
before. The relation between quasistaticity and reversibility leads us
to think that the $qR^{M\ot N}$ may be better suited to our purposes.

A single step sudden transformation is given analytically by
\begin{align}
 \label{adia_constr}
  R^{N-1\ot N}_{Ii} \;=\; \delta_{I,i} \frac{N-I}{N} + \delta_{I,i-1}
   \frac{I}{N} \;.
\end{align}\\
Iterating this relation it can be proved that the quasistatic
operators fulfill the recursion relation:
\begin{align}
 \label{adia_recurs}
  qR^{M\ot N}_{Ii} \;=\; \frac{M+1-I}{M+1} qR^{M+1\ot N}_{Ii} +
   \frac{I}{M+1} qR^{M+1\ot N}_{I+1,i} \;.
\end{align}\\
This relation allows to calculate the matrices using no matrix
products. This expression improves greatly the efficiency of the
numerical applications.

\begin{figure}[ht]
\vspace{0.4cm}
\centerline{
  \psfig{figure=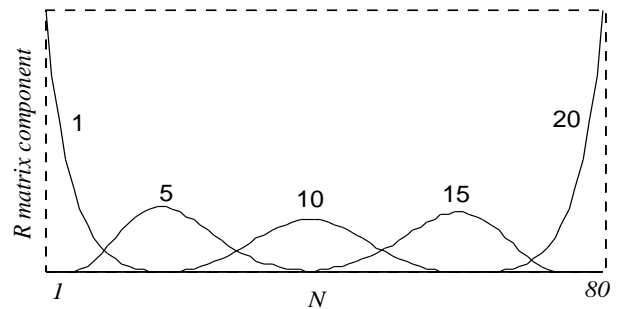,width=8.0cm,height=4.0cm}}
\vspace{0.2cm}
\caption{Some of the degrees of freedom which are retained by the
         quasi-static truncation operator proceeding from
         $80\to 20$ sites. Cells $1$, $5$, $10$, $15$ and $20$ are
         depicted. Notice that the ``cells'' are now overlapping
         and have slightly Gaussian nature.}
  \label{testqatm}
\end{figure}

The degrees of freedom which are retained by the quasistatic
truncation matrix are plotted in figure \ref{testqatm}. They are the
$E^N$ vectors given by the columns of $qR^{M\ot N}$.

%
%
Each of the discrete functions depicted in figure \ref{testqatm} may
be considered to represent a relevant degree of freedom when
truncating with the matrix $qR^{20\ot 80}$.  Although the functions
representing the degrees of freedom are now overlapping, they conserve
a true real-space nature. It should be noticed that the width of the
leftmost and rightmost cells is smaller than the one at the middle of
the interval. A consequence is the quite exact representation of the
boundary conditions.\\ 
It should be remarked that other authors have already introduced
overlapping blocks within RSRG applications~\cite{de00}. Inter-cell
correlations, which are the key to the most successful RSRG
algorithms~\cite{{wh98},{mrs96}}, are usually captured more easily
within an overlapping cells approach.

The most usual sub-discretization approach is the decimation method,
where one degree of freedom out of every $f$ is considered
relevant. This truncation scheme may not be represented within our
formalism. The reason is that the implementation on the field
discretization is given by the matrix
\begin{align}
 \label{deci_def}
   D_{Ii} \;=\; \delta_{fI,i} \;.
\end{align}
But the $R$ matrix (\ref{deci_def}) along with its SVD pseudo-inverse
yields a trivial dynamics, because the retained degrees of freedom are
{\em not in contact}. A possible solution to conserve linearity,
though losing the Moore-Penrose conditions (\ref{moore}).\\ 
A discrete Fourier Transform along with a cutoff might be a suitable
linear truncation procedure, but we shall not leave the RSRG setting:
our relevant degrees of freedom do have a local geometric meaning.

\section{Applications and Numerical Results}
\label{applnum}

This section discusses some numerical applications, both to linear and
non-linear examples.

\subsection{Heat Equation}
\label{secheat}

\begin{figure}
\psfig{figure=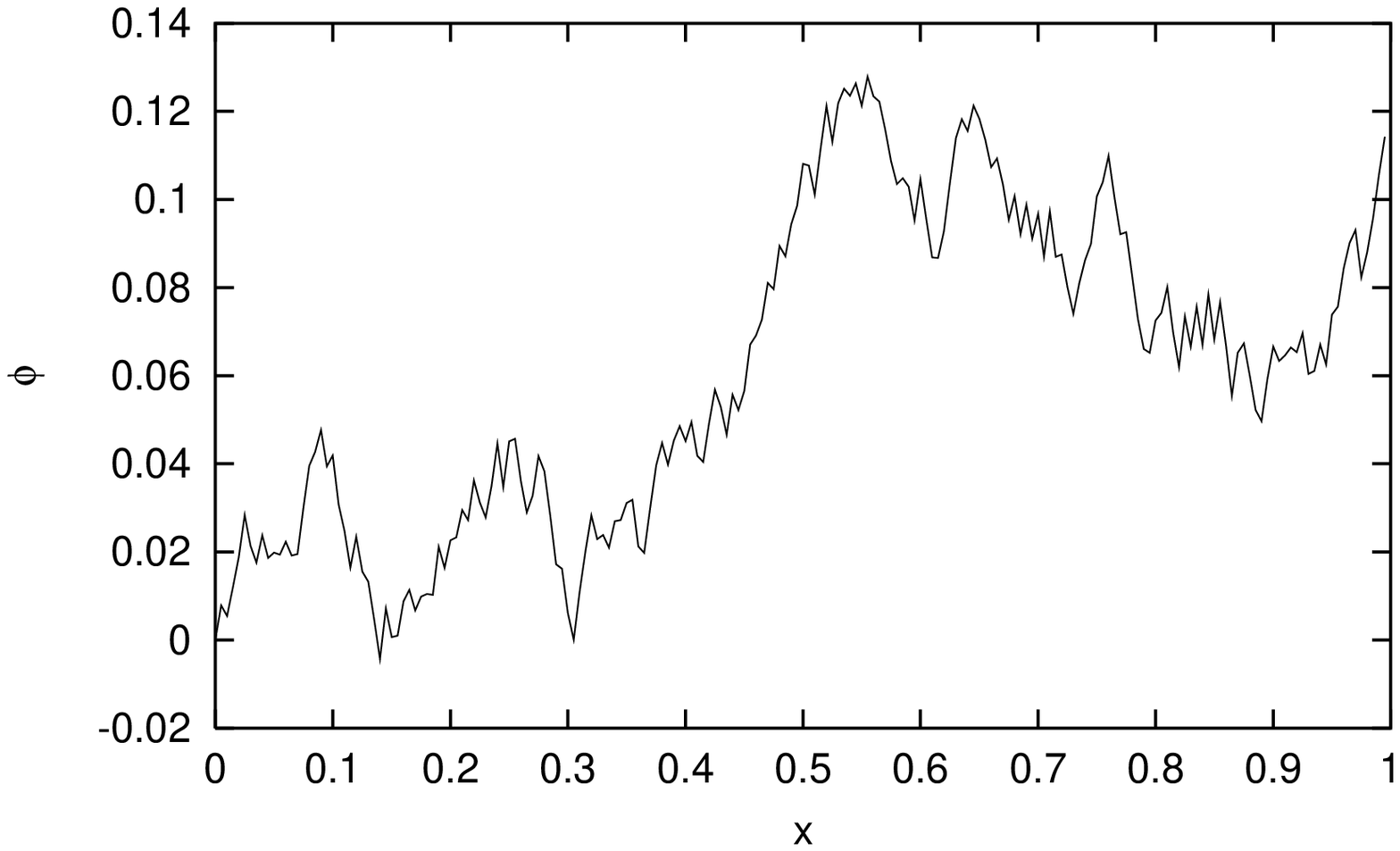,width=8.0cm}
\psfig{figure=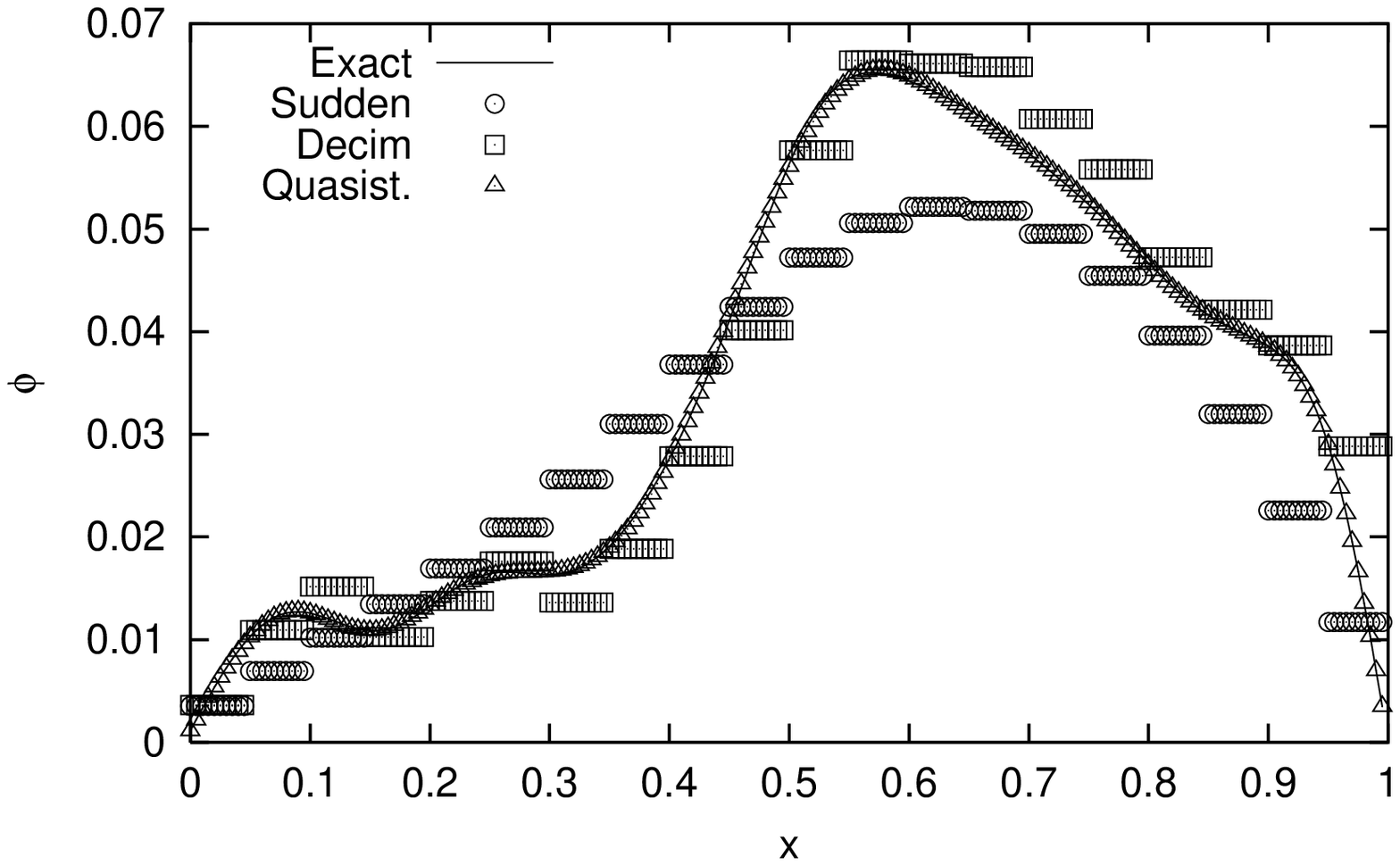,width=8.0cm}
\caption{An random increments function is taken as the initial
condition (up) with $200$ cells. Below, the continuous line shows the
exact evolution under the heat equation with $\kappa=1/2$, along $500$
time-steps with $\Delta t=5\cdot 10^{-6}$. The triangles are given by
the quasistatic approximation with $20$ degrees of freedom. The
circles represent the sudden approximation, and the squares follow the
sudden approximation, i.e. conventional symmetric coarse-graining.}
\label{heat_fig}
\end{figure}

The heat equation on any space is defined by stating that the
evolution operator is given by the minus the Laplacian on such a
space. It is known that the Laplacian operator may be sensibly defined
on a great variety of spaces~\cite{ro97}, including discrete
spaces~\cite{bo98}.\\
Our 1D interval shall always be $[0,1]$. As it is split into $N$
cells, the cells width is always $\Delta x=1/N$. The structure is
given by the discrete Laplacian matrix on a linear graph:
\begin{align}
 \label{heat_lapl}
   L_{ij} \;=\; 2\delta_{i,j}-\delta_{|i-j|,1} \;,
\end{align}
with fixed boundary conditions $L_{11}=L_{NN}=2$. The equation shall
be given by
\begin{align}
 \label{heat_eq}
  \pl_t{\bf \phi}_i \;=\; -\frac{\kappa}{\Delta x^2} L_{ij}\,{\bf \phi}_j \;.
\end{align}
The first test shall be a random increments initial condition, i.e. it
fulfills the equation:
\begin{align}
 \label{ran_incr_ini}
  {\bf \phi}_{i+1}\;=\;{\bf \phi}_i + r 
\end{align}
with $r$ a random variable with mean zero, equally distributed in an
interval of width $\Delta$. Using $N=200$, $M=20$ and $\Delta=1/4$ (a
quite severe reduction of a factor $10$) we obtain the results
depicted in figure \ref{heat_fig}.
%
%
The errors for the results of figure \ref{heat_fig} are summarized
in table \ref{heat_tab}.
%
%
\begin{table}[ht]
 \caption{Comparative of errors between different
          truncation schemes for heat equation on the random-increments
          initial condition depicted in figure \ref{heat_fig}.}
 \vspace{0.2cm}
 \begin{tabular}{l|c|c}
  \hline\hline
  Method & Real Space Error & Renormalized Space Error
  \\[0.1cm] \noalign{\hrule}
   Quasistatic & $\;0.53\%$ & $\;0.29\%$ \cr
   Sudden & $20\%$ & $19\%$ \cr
   Decimation & $13\%$ & $\;4.7\%$ \cr
  \hline\hline
 \end{tabular}
 \label{heat_tab}
\end{table}
Errors are noticed to be smaller in renormalized space. The reason is
that in real space two sources of error get mixed: the possibility of
representation of the initial data with the restricted degrees of
freedom and the dynamical relevance of the removed information. In
renormalized space only the second type of error contributes.\\ 
To examine the relevant scaling laws~\cite{ba96}, a discretization of
$\phi(x)=\delta(x-1/2)$ is defined on the $200$ cells partition, and
is normalized according to
\begin{align}
 \label{norma_cells}
   \sum_{i\;=\;1}^N \Delta x\, \phi_i \;=\; 1 \;.
\end{align}
Under time evolution, the peak becomes a Gaussian function and its
width $W$ follows the law 
\begin{align}
 \label{obs_diff}
  W(t) \;\equiv\; \sum_{i\;=\;1}^N i\Delta x\,\phi_i \sim t^{1/2} \;.
\end{align}
Equation (\ref{obs_diff}) can be proved to be exact also in the
discrete case as shown in the appendix.\\
Using the same constants as in the previous calculation, we have
performed a quasistatic simulation of the same problem, and depicted
in figure \ref{diff_obs} a log-log plot of the width against time:
%
%
\begin{figure}[ht]
\centerline{
  \psfig{figure=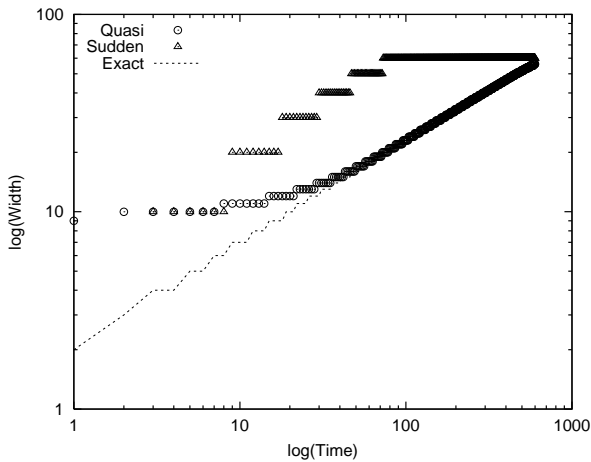,width=8.0cm}}
\caption{Log-log plot of the width of the Gaussian against time. The
steady straight line has slope $\approx 0.5$.}  
\label{diff_obs}
\end{figure}
The data from the quasistatic simulation in figure \ref{diff_obs} fit,
after a transient, to a straight line with slope $0.4990\pm
0.0001$. The exact field evolution yields exactly the same value,
without the transient. The sudden approximation saturates at long
times. Usual decimation gives a correct result.
%
\subsection{Low Energy States in Quantum Mechanics}
\label{seclowen}

Researchers in RSRG methods have spent many efforts in developing
techniques for the approximate obtention of the low-energy spectrum of
quantum mechanical problems~\cite{gms95}. The reason was not the
difficulty of the problem but of technical nature. With the development
of the Density Matrix RG, Correlated Blocks RG, etc. \cite{wh98,mrs96}
in the 90's, the problem was considered to be solved.\\ 
The quasistatic approach allows a very accurate approximation to the
lowest energies of many quantum mechanical 1D systems. The
transformation $H\to H'=RHR^\dagger$ may yield an effective
transformation of a Hamiltonian matrix, provided that the
transformation $R$ is orthogonal. In this case, the diagonalization of
$H'$ yields a {\em variational Ansatz} approach to the real
spectrum. The Ansatz is of the form
\begin{align}
 \label{ansatz_low}
   \left|\Psi\right>\;=\;\sum_{i=1}^M a_i \left|\phi_i\right>
\end{align}
where $\left|\phi_i\right>$ are the rows of $qR^{M\ot N}$ after a
orthonormalization procedure, and the $a_i$ are the variational
parameters.
The diagonalization of the quasistatically truncated Laplacian yields 
very precise values. For example, if $N=100$ and $M=10$, we obtain the
values for the spectrum of $-L$ exposed in table \ref{low_spec}.

%
%
\begin{table}[ht]
 \caption{Low energy spectrum of a particle in a box split into
          $100$ discrete cells, calculated through exact
          diagonalization, and two effective variational RG
          techniques: sudden and quasistatic transformations.}
 \vspace{0.2cm}
 \begin{tabular}{l|l|l|l|l|l}
  \hline\hline
   Method & & & & &
  \\[0.1cm] \noalign{\hrule}
   Exact & 0.000967435 & 0.0038688 & 0.008701 & 0.015460 & 0.02413 \cr
   Quasist. & 0.000967435 & 0.0038688 & 0.008701 & 0.015463 &
   0.02471 \cr
   Sudden & 0.008101410 & 0.0317493 & 0.069027 & 0.116917 &
   0.17153 \cr
  \hline\hline
 \end{tabular}
 \label{low_spec}
\end{table}
%
The bad results for the sudden approximation are a bit
misleading~\cite{term4}. For example, the real space error measured
according to the $L^2$ norm for the ground state is only around
$11\%$. The source of error is the lost of smoothness. The rest of
the eigenvalues (up to $10$) do not fit as well as the first ones.\\
The method has also been tested with the harmonic oscillator and other
potentials with equally good results, as long as the wave-functions
are smooth. In case of a potential given by $V_i=V(x_i)$, the
Hamiltonian operator is just $-L_{ij}+V_i\delta_{ij}$.

\subsection{Kardar-Parisi-Zhang Equation}
\label{secKPZ}

\begin{figure}[ht]
\vbox{
\psfig{figure=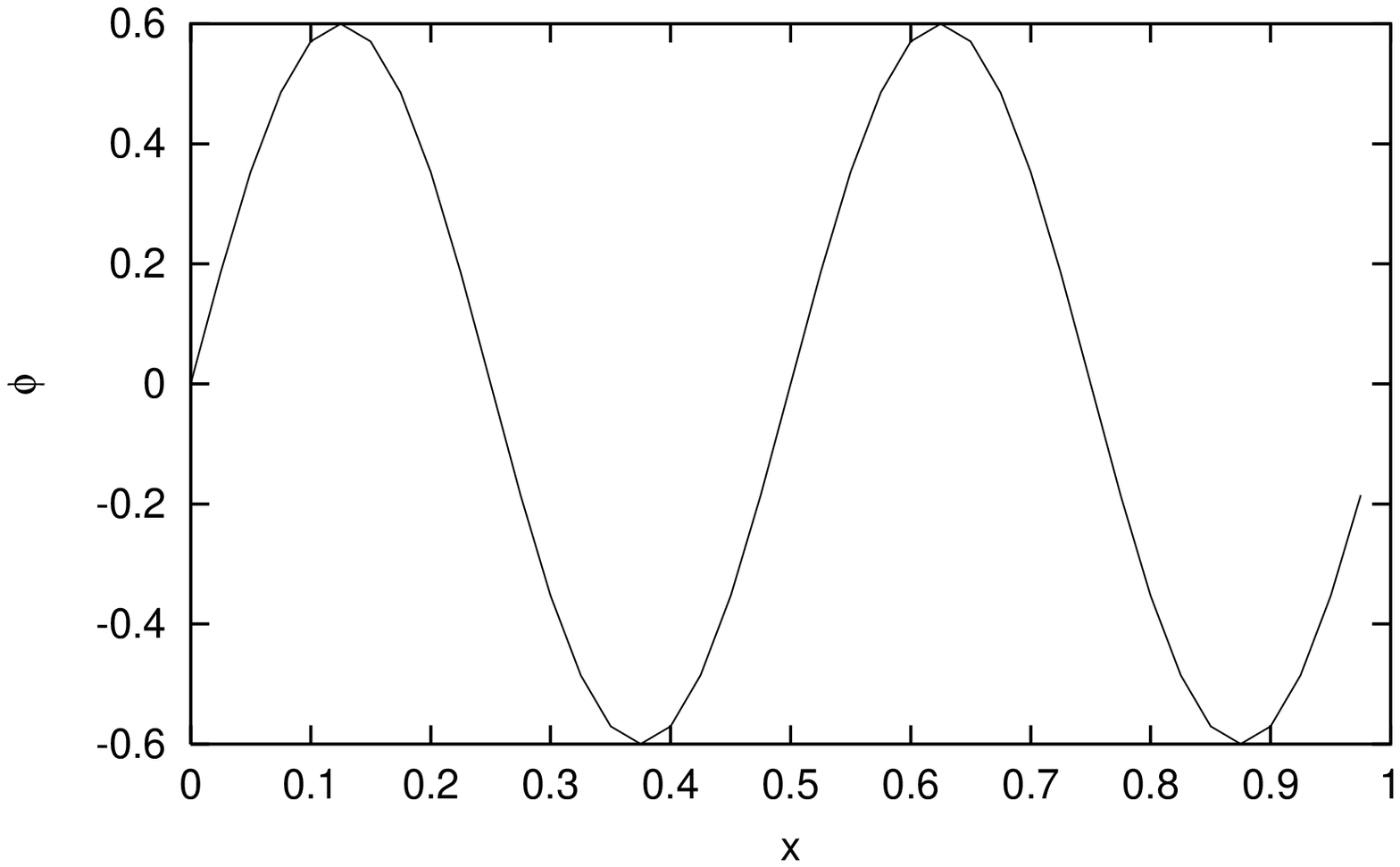,width=8cm}
\psfig{figure=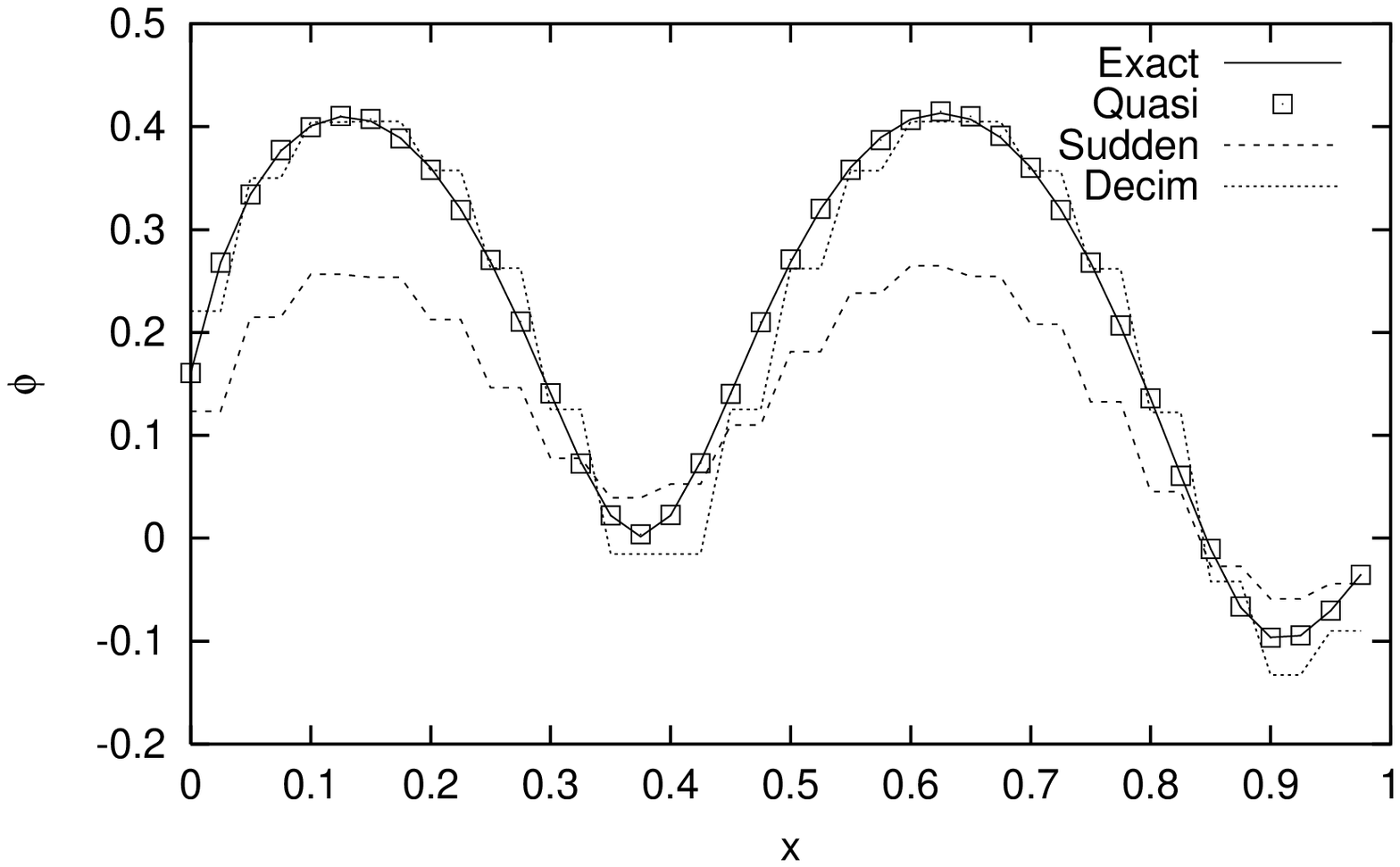,width=8cm}
}
\caption{A sinusoidal surface profile evolved by the KPZ dynamics with
the parameters explicitly given in the text. Notice that a slight
asymmetry in the initial function (a lattice artifact) develops a high
asymmetry in the exact and quasistatic approximations.}
\label{KPZ_fig}
\end{figure}

\begin{figure}[ht]
\vbox{
  \psfig{figure=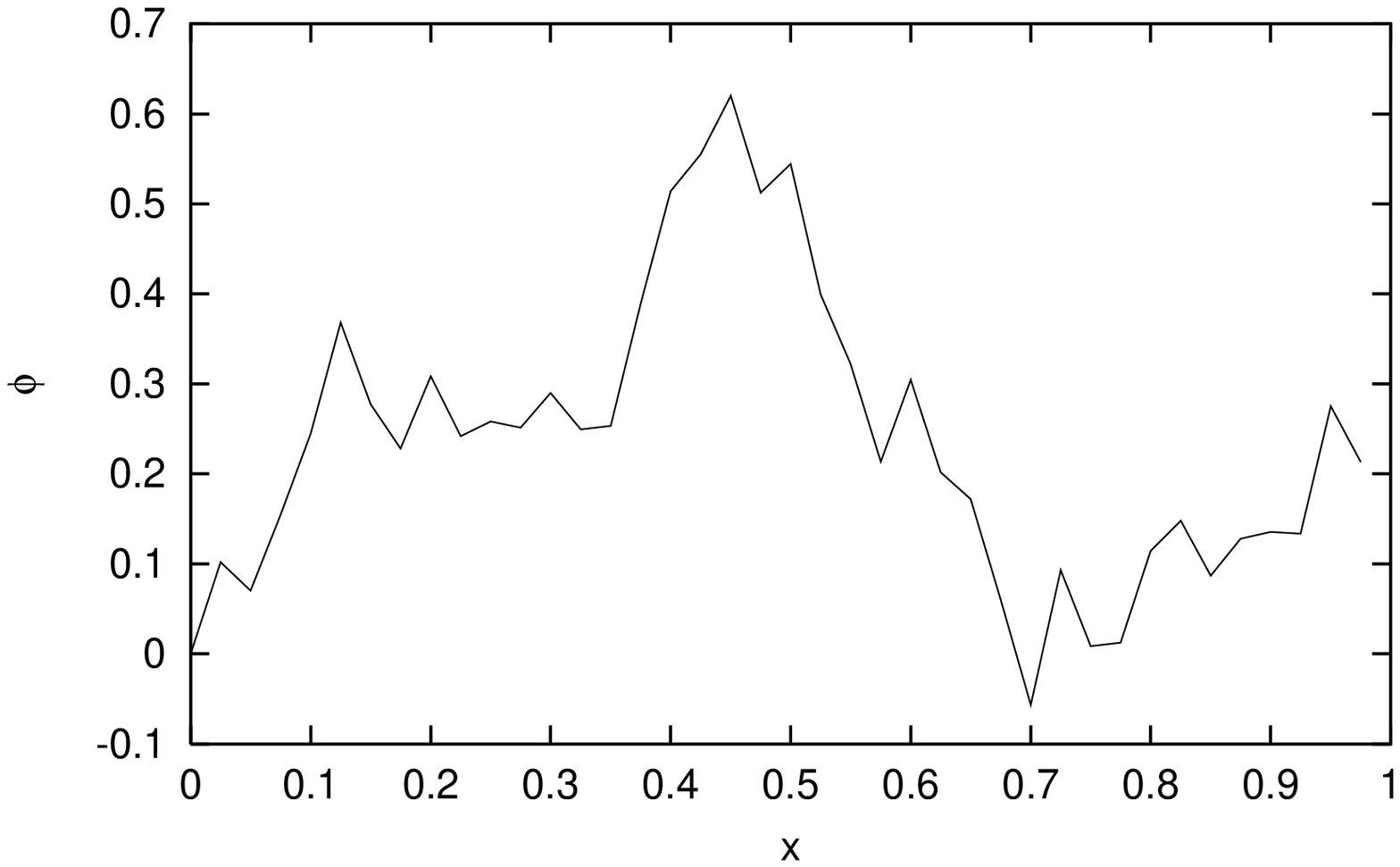,width=8cm}
  \psfig{figure=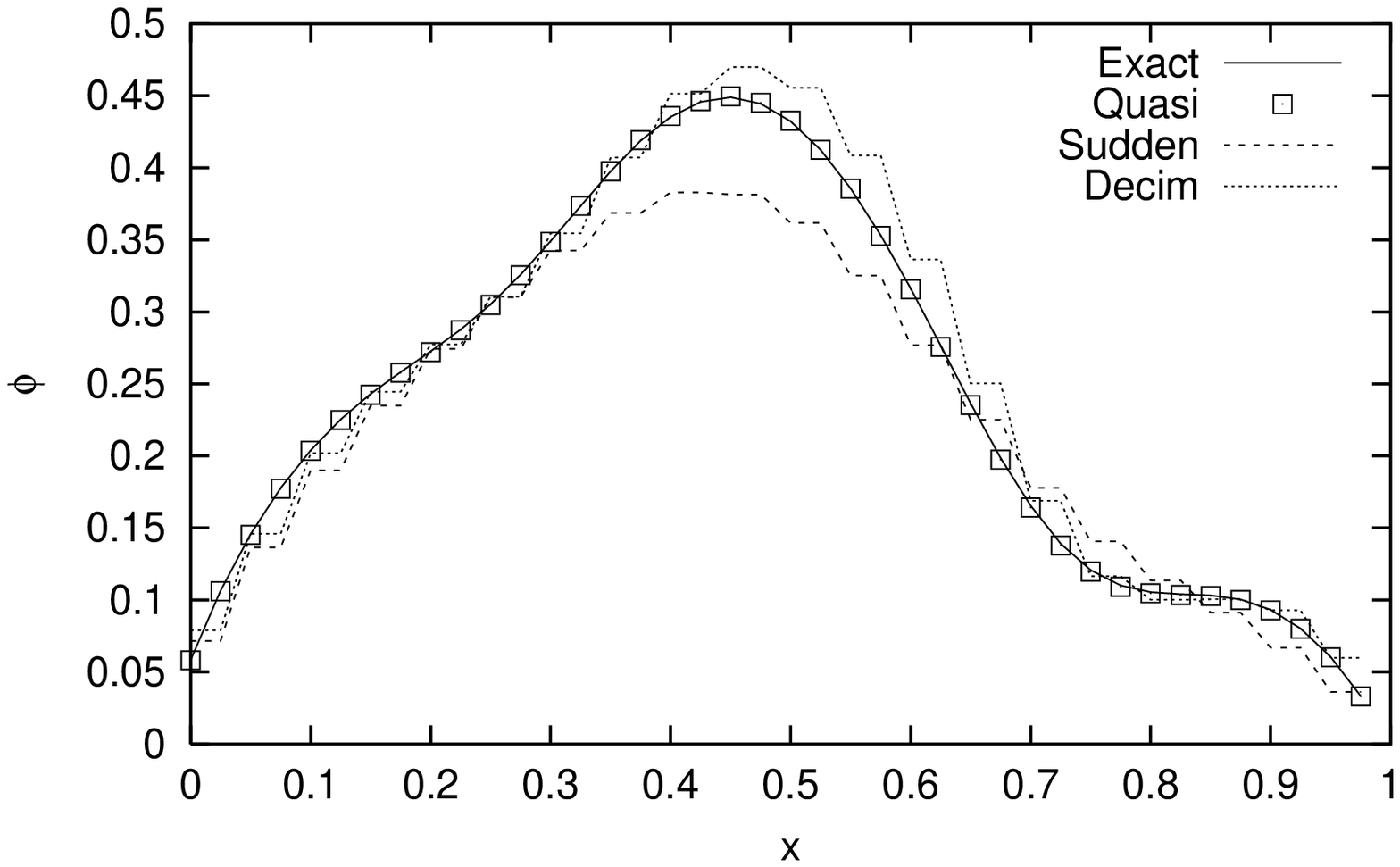,width=8cm}}
\caption{Random increments function (above) built in the same way as
that of figure \ref{heat_fig}. Below, the exact evolution is
displayed by the continuous line. Squares mark the quasistatic
truncation approximation, while the dashed lines follow the sudden and
the decimation truncations.}  
\label{KPZ_rand_fig}
\end{figure}

The Kardar-Parisi-Zhang (KPZ) equation is widely used as a model
of stochastic and deterministic surface growth~\cite{kpz86}.
Here we use the deterministic form defined as
\begin{align}
 \label{KPZ_eq}
  \pl_t\phi\;=\;\lambda|\nabla\phi|^2 + \kappa\nabla^2\phi
\end{align}
representing a surface in which absorption/desorption phenomena take
place. The squared gradient term shall be implemented through the
quadratic operator
\begin{align}
 \label{KPZ_squarred}
  K_{ijk}\;=\;\frac{1}{4}\left( \delta_{j,i+1}-\delta_{j,i-1}\right) 
               \left(\delta_{k,i+1}-\delta_{k,i-1}\right)
\end{align}
which is obtained through the centered derivatives approximation to
the gradient~\cite{ptvf92}. Boundary conditions are imposed for
which forward and backward derivatives are employed.\\
The first test evolves an initial condition given by a sinusoidal
function $\phi(x)=\sin(4\pi x)$ with $x\in[0,1]$. The resolution
change is $40\to 20$ and $2000$ time steps with $\Delta t=5\cdot
10^{-6}$ were simulated.  Figure \ref{KPZ_fig} shows the results for
$\lambda=2$ and $\kappa=1/2$.
%
%
The errors for such a test are given in table \ref{KPZ_sin_tab}.
%
%
\begin{table}[ht]
 \caption{Errors in the evolution of a sinusoidal initial condition
          under KPZ equation, corresponding to the results of figure
          \ref{KPZ_fig}. The parameters are explicitly given in the
          text.}  
\vspace{0.2cm} 
\begin{tabular}{l|c|c} \hline\hline
   Method & Real Space Error & Renormalized Space Error\\[0.1cm] 
   \noalign{\hrule} 
   Quasist. & $\;0.5\%$ & $\;0.2\%$ \cr 
   Sudden & $39\%$ & $38\%$ \cr 
   Decim. & $15\%$ & $\;8\%$ \cr \hline\hline 
\end{tabular} \label{KPZ_sin_tab}
\end{table}
A different test was carried out with a random increments function,
as for the heat equation. The rest of the parameters are the same
as in the previous simulation. The results of this simulation are
displayed in figure \ref{KPZ_rand_fig} and the numerical errors are
provided in table \ref{kpz_rand_tab}.
%
%
%
%
\begin{table}[ht]
\caption{Errors corresponding to the field evolution of a
random-increments initial condition, shown in
figure \ref{KPZ_rand_fig}.}
\vspace{0.2cm}
\begin{tabular}{l|c|c}
\hline\hline
Method & Real Space Error & Renormalized Space
Error\\[0.1cm] 
\noalign{\hrule}
   Quasist. & $\;0.23\%$ & $\;0.23\%$ \cr
   Sudden & $12\%$ & $11\%$ \cr
   Decim. & $\;9.4\%$ & $\;6.5\%$ \cr
\hline\hline
\end{tabular}
\label{kpz_rand_tab}
\end{table}

Some more nonlinear equations have been tried, such as Burgers
\cite{term5} and others, with comparable results. We encourage the
reader to experiment.

\subsection{Efficiency issues.}

A problem which must be remarked is that the approximation renders new
evolution generators which may have a greater number of non-null
entries than the originals. The elements typically decrease in
magnitude as a power of their distance to the diagonal, albeit they
often alternate signs. This corresponds to the non-local space-time
effects remarked by Goldenfeld et al. \cite{hgk01}.\\
This fact forces the practitioner to make computational complexity
estimates before trying this method. Various factors should be
pondered:

$\bullet$ {\em Reduction factor attainable for a given
equation}. KPZ stands more than $50\%$ reduction for a wide set of
initial conditions. The heat equations stands more than $90\%$.

$\bullet$ {\em Availability and stability of local explicit
methods}. If implicit methods must be used, or the equation has
non-local nature, then the original equation is already long ranged
and no loss of efficiency comes from applying the RSRG recipe
described.

\section{Conclusions and Future Prospectives}
\label{secfuture}

A new formalism has been provided to deal with the reduction of
degrees of freedom for a wide set of field evolution equations. The
basis of the formalism is the {\sl integral} specification of the
field values (i.e.: it is related to finite volume methods). The key
concept to find the transformation between a partition of space and
another is the {\sl overlapping} of cells.

Our specific recipe stands removal of $90\%$ of the degrees of freedom
without distortion for linear PDE such as the heat equation, and
$50\%$ reduction without appreciable loss of accuracy for KPZ and
related nonlinearities.

The main handicap of the technique is shared by all known strategies
to the reduction of degrees of freedom: the appearance of nonlocal
effects which may spoil the efficiency \cite{hgk01}. Future works on
this algorithm should try to find suitable short-ranged approximations
to the renormalized evolution generators. Also the extension to
stochastic PDE makes nonlocal effects appear: a spatially white noise
shall develop a nontrivial covariance matrix. The eigenfunctions of
this matrix would be the appropriate basis.

It is easy to generalize the formalism to higher dimensions, but the
algorithms to find cell overlappings is trickier. Nevertheless, fields
of vectorial nature do not fit well in this formalism. The authors are
developing a ``difference forms'' theoretical frame to deal with them,
in the line traced by Katz and Wiese~\cite{kw97}.

But the main interest of the authors at the present moment is a
different extension: to find an algorithm in which the degrees of
freedom are not of geometric nature, but are {\em chosen} by the
equation itself.


\begin{acknowledgments}

The authors would like to thank R.~Cuerno, M.A.~Mar\-t\'{\i}n-Del\-gado,
S.N.~Santalla and G.~Sierra for very useful discussions.

\end{acknowledgments}

\appendix
\section{Scaling of the discrete heat equation.}

In this appendix the exactness of relation (\ref{obs_diff}) subject to
any coarse-graining procedure keeping the normalization condition
(\ref{lin_est}) is proved.\\
We generalize the definition in (\ref{obs_diff}) to the expectation
value for any observable $\cal O$ on a one dimensional lattice
composed of $N$ sites according to
\begin{align}
 \label{exp_diff}
 \braket{\cal O}_t\;\equiv\;\sum_{i=1}^{N}\,{\cal O}_{t,i}
                          \,\Delta x\, {\bf \phi}_{t,i}
\end{align}
with the total time $t\,=\, n\, \Delta t$ and $n$ the number of
discrete time evolutions.\\
Relation (\ref{obs_diff}) also describes Brownian motion on a 1D
lattice. According to Wick's theorem~\cite{id94}, it is sufficient to
prove the linear dependence of the second moment $\braket{x^2}$ on
time for any discretization scale, as the following proposition
states:\\
{\bf Proposition.} The second moment $\braket{x^2}$ as defined by
definition (\ref{obs_diff}) for the diffusion field $\phi$ is given
by (supposing free or periodic boundary conditions):
\begin{align}
 \label{theme}
  \braket{x^2}_t\;=\;2 \kappa\,t\,+\,C(\phi_{t=0})\;,
\end{align}
subject to the normalization constraint (\ref{norma_cells}). Here,
$C(\phi_{t=0})$ is a constant which depends on the initial field
configuration and, for a $\delta$ initial condition, $C(\phi_{t=0})=0$.
In equation (\ref{theme}), $t$ is the time, $\kappa$ is the diffusion
constant and no dependence on the discretization scale $\Delta x$ is
involved.

Using definition (\ref{exp_diff}) to define the second moment
$\braket{x^2}$ we have

\begin{widetext}
\begin{align}
 \label{theme2}
 \braket{x^2}_{t+1}
  = \sum_{i=1}^{N}\,\Delta x\,(i\,\Delta x)^2\,{\bf \phi}_{t+1,i}
    \;=\; \sum_{i=1}^{N} i^2\,(\Delta x)^3
     \left[\frac{\Delta t \cdot\kappa}{(\Delta x)^2}
      \left({\bf \phi}_{t,i-1} - 2\cdot {\bf \phi}_{t,i}
        + {\bf \phi}_{t,i+1}\right) \,+\, {\bf \phi}_{t,i} \right]\,.
\end{align}
\end{widetext}

The evolution equation uses a discrete Laplacian ({\ref{heat_lapl})
and a forward time Euler scheme \cite{ptvf92}. Some algebra and index
shifting, along with the supposition of either free or periodic
boundary conditions lead to

\begin{align}
 \label{theme6}  
  \braket{x^2}_{t+1} \;=\; 2 \kappa\,{\Delta t}\,\sum_{i=1}^{N}
      \Bigl\{\, \Delta x \,{\bf \phi}_{t,i}\Bigr\}
    + \braket{x^2}_{t}
\end{align}

The equation is rewritten, taking into account that
(\ref{norma_cells}) is valid for all time, as
\begin{align}
 \label{theme7}  
  \braket{x^2}_{t+1} \;=\; 2\kappa\,{\Delta t} \, +\, \braket{x^2}_{t}\;.
\end{align}
Iterating the procedure $n$ times yields the final result
\begin{align}
 \label{theme8}  
  \braket{x^2}_{t+1} \;=\; 2 \kappa\,\left(t+1\right)
                          \, +\, \braket{x^2}_{t=0}\;.
\end{align}
Defining $C(\phi_{t=0})\,\equiv\,\braket{x^2}_{t=0}$ and changing the
index $t+1$ to $t$ we get the result stated in the above proposition.
If the initial field configuration $\phi_{t=0}$ is provided by the
$\delta$ peak which was used to generate figure \ref{diff_obs}
equation (\ref{theme8}) simplifies to
\begin{align}
 \label{theme9}  
  \braket{x^2}_{t} \;=\; 2 \kappa\,{t} \;.
\end{align}
Equation (\ref{theme9}) is equivalent to the calculation of the
mean squared distance of a random walker after the time $t$
starting at the center position, i.e. the location of the $\delta$ peak.
%



\begin{thebibliography}{99}
\bibitem{term2} The perfect actions project aims at the removal of
lattice artifacts on the discretizations of a continuous system. This
work shall assume the discrete equation to be correct, along with all
its lattice-dependent results.
\bibitem{term5} Of course, Burgers is intimately related to KPZ, but
this does not imply that an algorithm which suits well to one of them
must suit also the other.
\end{thebibliography}
\end{document}